\documentclass{article}
\usepackage{spconf,amsmath,graphicx}
\usepackage{multirow}
\usepackage{booktabs} 

\title{Perceptual-based Deep-Learning Denoiser as a Defense Against Adversarial Attacks on ASR Systems}
\name{Anirudh Sreeram, Nicholas Mehlman, Raghuveer Peri, Dillon Knox, Shrikanth Narayanan}
\address{Signal Analysis and Interpretation Laboratory (SAIL),\\
Ming Hsieh Department of Electrical and Computer Engineering, Viterbi school of Engineering, \\University of Southern California\\
E-mail-{\{asreeram, nmehlman, rperi, dillonkn, shrikann\}@usc.edu}}
\begin{document}
\topmargin=0mm

\maketitle
\begin{abstract}
In this paper we investigate speech denoising as a defense against adversarial attacks on automatic speech recognition (ASR) systems. Adversarial attacks attempt to force mis-classification by adding small perturbations to the original speech signal. We propose to counteract this by employing a neural-network based denoiser as a pre-processor in the ASR pipeline. The denoiser is independent of the downstream ASR model, and thus can be rapidly deployed in existing systems. We found that training the denoisier using a perceptually-motivated loss function resulted in increased adversarial robustness without compromising ASR performance on benign samples. Our defense was evaluated (as a part of the DARPA GARD program) on the 'Kenansville' attack strategy across a range of attack strengths and speech samples. An average improvement in Word Error Rate (WER) of about $7.7\%$ was observed over the undefended model at 20 dB signal-to-noise-ratio (SNR) attack strength.

\end{abstract}
\begin{keywords}
Kenansville, denoiser, pre-processor, adversarial attack, black-box attack
\end{keywords}
\section{Introduction}
\label{sec:intro}
Automatic speech recognition (ASR) systems have become integrated in many aspects of our daily lives. From our cars and our kitchens and living rooms to our mobile phones, we have come to expect that much of our interaction with technology will be voice-mediated. While this level of adoption may be highly convenient, it also introduces a number of security concerns. In particular, these systems are vulnerable to adversarial attacks in which a malicious actor can add small perturbations to the speech signal in an attempt to force the ASR to mis-transcribe the utterance \cite{imperceptible}. Because the perturbations are generally imperceptible to a human listener, they can usually pass as unaltered/undetected speech samples, and thus manipulate the ASR system without the user's knowledge \cite{zelasko2021adversarial}. There are two primary classes of adversarial attacks. In "white-box" attacks, the inner workings (i.e., the gradients) of the ASR model are known to the attacker \cite{Jati_2021}. This knowledge is then used in constructing the adversarial perturbations  \cite{Jati_2021}.  "Black-box" attacks, on the other hand, must be accomplished without any special knowledge of the ASR model \cite{Jati_2021}. Both the black-box and white-box attacks can be deemed as external interference or noise being added to the clean speech signal. Therefore, a pre-processor that can remove or mitigate  these noises can act as a defense against these attacks; the premise of this paper. 

\par In this paper, a pre-processor which is a neural network based denoiser, is introduced as a defense against the 'Kenansville' black-box attack presented in \cite{kenansville}. The Kenansville attack employs a Discrete Fourier Transform (DFT) to first decompose the speech signal into its spectral components and all of the components that fall bellow a certain amplitude threshold are discarded \cite{kenansville}. The Kenansville attack is an un-targeted attack, since it does not seek to produce a specific erroneous transcription \cite{Jati_2021}\cite{kenansville}. Our denoiser decreased adversarial word error rates by approximately $7.7\%$ when applied to the DeepSpeech ASR model available through Armory as a part of the DARPA Guaranteeing AI Robustness against Deception (GARD) program \footnote{https://github.com/twosixlabs/armory} \cite{deepspeech}. The proposed method tries to counter the effects of the attack by removing the noise and making the denoised output perceptually similar to the benign  speech signal.

The rest of the paper is organized as follows: The related prior work is presented in Section $2$, in Section $3$ we discuss the proposed defense, followed by the experiments in Section $4$, results and discussion in Section $5$, conclusion in Section $6$ and References in Section $7$.\par

\section{Related prior work}
Several methods have been proposed to defend ASR systems against adversarial attacks. The most straightforward is adversarial training, which involves including adversarial samples in the data set used to train the speaker-recognition model \cite{Jati_2021}. Although this may improve the system's robustness under attack, it also has the drawback of significantly degrading performance on benign samples \cite{Jati_2021}. In \cite{kenansville}, the authors evaluate the Kenansville attack against an adversarial training defense. For a $20$ dB SNR attack, they showed that this defense decreased the error-rate by approximately $36\%$ relative to the undefended model \cite{kenansville}. However, it also nearly doubled the error rate for benign speech samples \cite{kenansville}. 
\par
Another defense method known as randomized smoothing adds white Gaussian noise to the speech signal prior to the input of the ASR model \cite{zelasko2021adversarial}. The noise is intended to interfere with the carefully crafted adversarial perturbations \cite{zelasko2021adversarial}. However, here again, the addition of noise presents obvious challenges to the model's accuracy on benign samples \cite{zelasko2021adversarial}.
\par
Adversarial Lipschitz Regularization (ALR) introduces an additional loss function to the ASR training procedure based on the Lipschitz smoothness measure \cite{Jati_2021}. This loss incentivizes the model to learn a smooth mapping, in which small changes to the input do not produce drastically different predictions \cite{Jati_2021}. This makes it more difficult for adversarial actors, since they must add larger perturbations in order to achieve mis-classification \cite{Jati_2021}.
\par
Finally, MP3 compression has also been tested as a defense under the rationale that it removes imperceptible perturbations in the speech signal \cite{mp3}. As with randomized smoothing, the compression is applied as a pre-processing stage in the ASR pipeline \cite{mp3}. However, in our evaluations we found that this method produced poor results on both adversarial and benign samples, suggesting that ASR systems are not robust to MP3 compression as shown in previous works \cite{borsky2015advanced}.

\section{PROPOSED DEFENSE}
\label{sec:defense}

\begin{figure}[htb]
\centerline{\includegraphics[width=8.5cm]{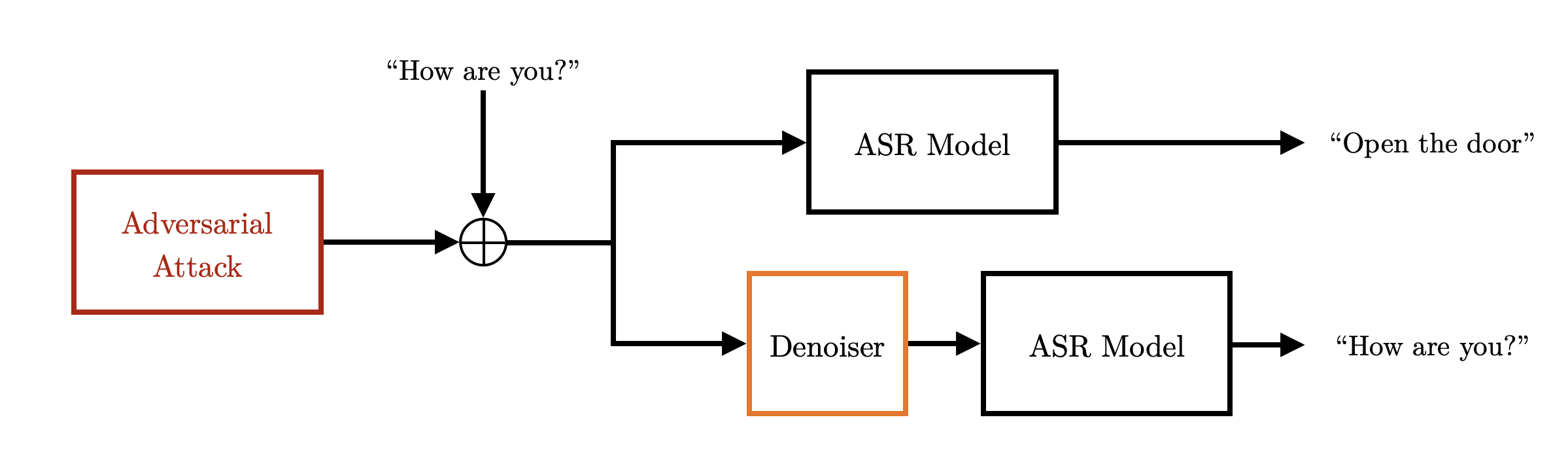}}
\caption{Adversarial attack against the defended and undefended ASR system.}
\label{fig:defense}
\end{figure}

To defend against attacks, we introduce a denoiser designed to counteract adversarial perturbations prior to processing with the ASR model. Figure \ref{fig:defense} shows the  schematic block diagram of the proposed defense against the adversarial attacks. The adversarial attacks are added to the original speech signal, leading to an undesirable output if not defended. The denoiser, which is completely differentiable, functions as a pre-processor to the existing ASR system. Unlike other defense methods (i.e., adversarial training and ALR, our approach is model-agnostic. This means that it can be added to existing ASR systems without the need for retraining, a process that can be expensive and time consuming.

Inspired by the success in speech enhancement achieved by \cite{denoiser}, we adopted the model presented by the authors in \cite{denoiser} as the pre-processing defense which serves as the pre-trained denoiser. The denoiser uses the DEMUCS architecture, which is an encoder-decoder based deep neural network with U-Net skip connections \cite{ronneberger2015u} and a sequence modeling network at the end of the encoder stage. We refer the readers to \cite{denoiser} for a more detailed explanation of the denoiser model.

An L1 loss applied in the waveform domain and a multi-resolution loss applied to the magnitude of the short-time Fourier Transform (STFT) are employed to jointly optimize the time and frequency components of the speech sample.

\begin{equation}
    loss_{L1}(y,\hat{y}) = \| y-\hat{y}\|_1
\end{equation}
where $y$ and $\hat{y}$ are the clean and denoised signal respectively and $\| .\|_1$ is the L1 norm. 
\begin{equation}
    l_{stft}(y,\hat{y}) = L_{sc}(y,\hat{y}) + L_{mag}(y,\hat{y})
\end{equation}
The STFT loss is defined as the sum of the spectral convergence ($L_{sc}$) loss and the magnitude ($L_{mag}$) loss. 

\begin{equation}
    L_{sc}(y,\hat{y}) = \frac{\| |STFT(y)|-|STFT(\hat{y})| \|_F}{\| |STFT(y)| \|_F}
\end{equation}

\begin{equation}
    L_{mag}(y,\hat{y}) = \frac{1}{T}~\|\log(|STFT(y)|) - \log(|STFT(\hat{y}|)\|_{1}
\end{equation}

where $\|.\|_{F}$ and  $\|.\|_{1}$ are the Frobenius and the L1 norms, respectively. The multi-resolution STFT is defined as the sum of all STFT loss functions using different STFT parameters.

\begin{equation}
    loss_{stft}(y,\hat{y}) = \sum_{i=1}^{M} l_{stft}^{(i)}(y,\hat{y}) 
\end{equation}
where $M$ is the number of STFT losses, and each $l_{stft}^{(i)}$ applies 
the STFT loss at a different resolution with number of FFT bins $\in {\{512, 1024, 2048\}}$, hop sizes $\in {\{50, 120, 240\}}$, and lastly window lengths $\in {\{240, 600, 1200\}}$.

In addition to the above mentioned optimization, in order to build a more robust defense against the attacks, we incorporate a perception-based distance metric to the above losses to fine-tune the pre-trained denoiser. We employ a perceptually motivated loss, which tries to minimize the perceived difference between the clean and denoised signals. The perceptual loss is derived from a neural network based perceptual distance metric which is trained with the objective of finding the similarity between two speech signals by using just noticeable difference (JND) thresholding and human judgment as targets. A high quality perceptual distance metric $D$ would provide a small distance $D(y, \hat{y})$ if human judges feel they are the same recording, and a larger distance if they are judged to be different \cite{manocha2020differentiable} 

\begin{equation}
    D(y,\hat{y}) = \|F(y) - F(\hat{y})\|_1
\end{equation}

where $F$ is the activation of the deep network embedding \cite{manocha2020differentiable}. The overall loss to be optimized is the summation of the $L1$ loss, the multi-resolution STFT loss and the perceptually inspired loss. By optimizing the three losses together, we not only try to remove the noise but also make it perceptually closer to the benign speech signal, thereby making it difficult to generate attacks that are not similar to the speech signals. The pre-trained model \cite{manocha2020differentiable} was used to obtain the perceptual distance, and we do not update the distance metric model during our training. The distance metric obtained is only used in the loss term to update the denoiser model weights.

\begin{equation}
    loss_p = \alpha * loss_{L1}(y,\hat{y}) + \beta * loss_{stft}(y,\hat{y}) + \gamma *D(y,\hat{y})
    \label{eq:lossp}
\end{equation}
The parameters $\alpha$, $\beta$ and $\gamma$ control the effect of each of the $3$ losses. Experiments were performed to tune these parameters in order to optimize the effect of each of the losses. 

We load the weights for the denoiser from the publicly available pre-trained denoiser model. This model is then fine-tuned with the objective of minimizing the overall cost presented in \eqref{eq:lossp}, thereby helping the denoiser to not only reproduce cleaner speech signals but also perceptually better ones; hence, making it robust to attacks that are imperceptible to humans.  

\section{EXPERIMENTS}
\label{sec:experiments}
We evaluated the performance of our preprocessor defense against the Kenansville attack at a range of attack SNR values. The defense was also tested on benign samples (i.e. with no adversarial perturbations) to ensure that the preprocessor did not substantially degrade the baseline accuracy of the ASR system. All the experiments for the denoiser were performed using PyTorch framework. The attack samples and ASR evaluations were performed using the armory toolbox available through Armory \cite{deepspeech}. 

\subsection{Data} 
For training the denoiser model, we used a subset of the Valentini data \cite{Valentini}, which contains about $11572$ and $824$ clean speech recordings from $28$ speakers, for training and testing respectively, all sampled at $48$kHz. Additive White Gaussian noise and pink noise were added to the clean files of the training data at SNRs of $18$, $21$, $24$, $27$ and $30$dB randomly to get the noisy speech recordings.

For evaluating the robustness of denoiser as a defense, the Librispeech dataset \cite{librispeech} was used. Specifically we used the "test clean" set, which contains around $5$ hours of labeled speech derived from audio books \cite{librispeech}. Approximately $40$ different speakers are included in the corpus \cite{librispeech}.

\subsection{Training}
The DEMUCS model based denoiser that was trained on the Valentini \cite{Valentini} and the DNS \cite{dns} dataset for 400 epochs served as the pretrained off-the-shelf denoiser. This model was optimized with only the $L_{1}$ and the STFT losses. For fine-tuning this model, we used the noise augmented Valentini dataset as training data with the corresponding clean data as the target and optimized on a combination of the $L_{1}$, STFT and perceptual based loss \ref{eq:lossp}. The pre-trained model was fine-tuned for $10$ epochs with a learning rate of 3e-5.

Multiple experiments were performed in fine-tuning the $\alpha$, $\beta$ and $\gamma$ parameters. The setup of $0.45$, $0.45$ and $0.45$ for $\alpha$, $\beta$ and $\gamma$ respectively, were found to produce the best results in defending adversarial attacks.

\subsection{ASR}

\par In our experiments, we used the DeepSpeech ASR model available through Armory \cite{deepspeech}. The DeepSpeech model, presented in \cite{deepspeech} is an end-to-end deep neural network. The model operates on the spectrogram of the input signal which is fed through a series of convolutional layers \cite{deepspeech}. These are followed by a sequence of recurrent layers, and a final fully-connected layer that produces the output \cite{deepspeech}.
The ASR model uses pre-trained weights and was kept unaltered throughout our evaluations.

\begin{figure}[htb]
\centerline{\includegraphics[width=8.5cm]{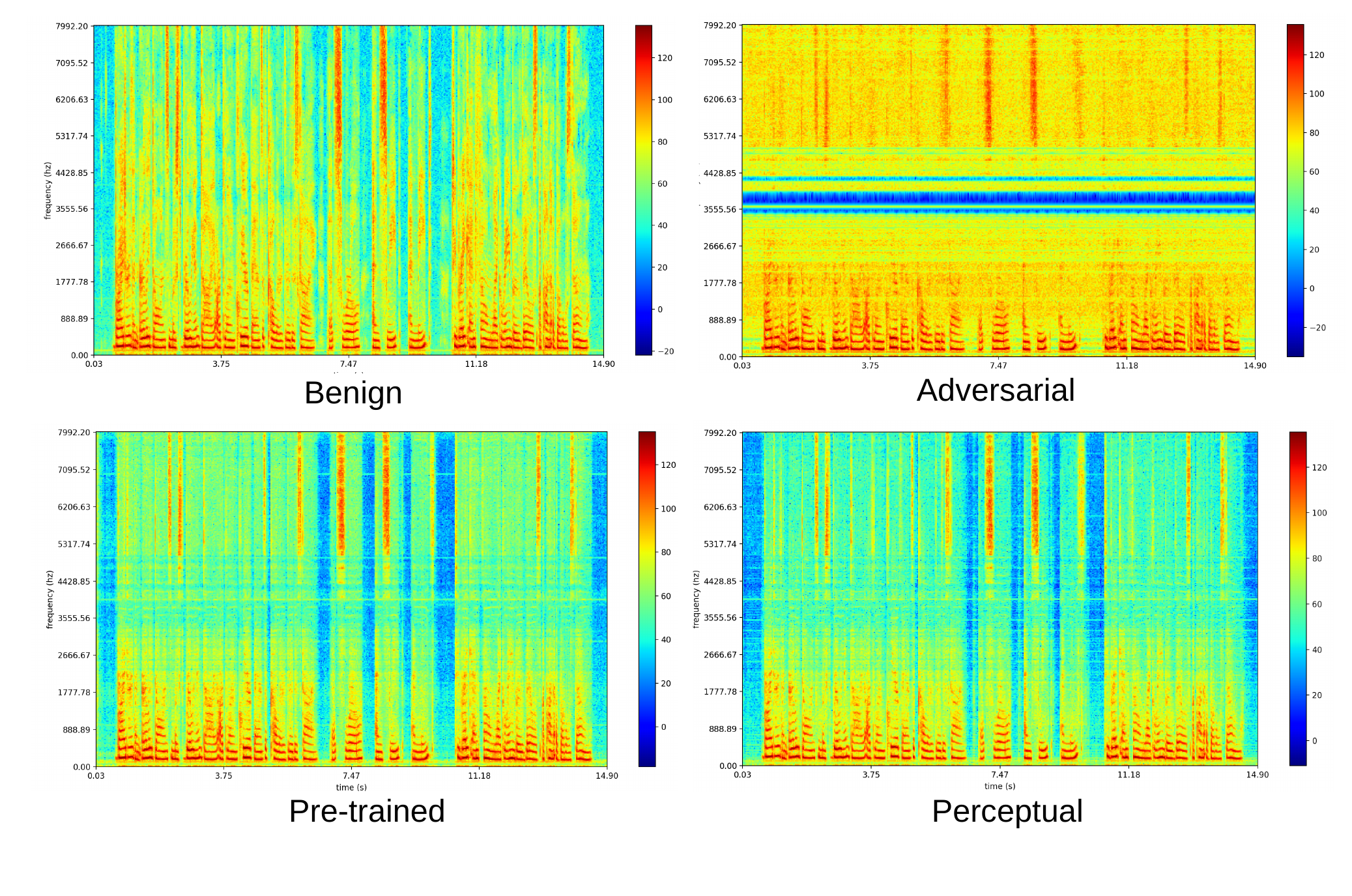}}
\vspace{-0.5cm}
\caption{Spectrograms of the Benign, Kenansville attacked speech signal (20 dB), Pre-trained denoiser and Perceptual loss based denoiser.}
\label{fig:spectrogram}
\end{figure}

\section{Results and discussion}
\label{sec:typestyle}

\begin{table}[]
\centering
\resizebox{\columnwidth}{!}{
\begin{tabular}{c|c|c|c|c|c|c}
\toprule
\multirow{2}{*}{Defense} & \multicolumn{5}{c|}{SNR (dB)}                           & \multirow{2}{*}{Benign} \\ \cline{2-6}
                         & 10             & 15             & 20    & 25    & 30    &                         \\ \midrule
Undefended               & 83.38          & 56.53          & 27.69 & 14.81 & 11.35 & \textbf{9.92}                    \\
MP3                      & 85.19          & 62.59          & 36.22 & 22.74 & 17.08 & 13.08                   \\
Pre-trained              & \textbf{70.42}          & 43.46          & 21.73 & 13.81 & 11.17 & 9.99                    \\
Perceptual               & 71.31 & \textbf{42.04} & \textbf{19.93} & \textbf{12.59} & \textbf{10.82} & 10.03                   \\ \bottomrule
\end{tabular}
}
\caption{Word Error Rates (\%) for various defenses against the `Kenansville' attack strategy across a range of attack strengths.}
\vspace{-0.5cm}
\label{table:2}
\end{table}

\begin{figure}[htb]
\centerline{\includegraphics[width=8.5cm]{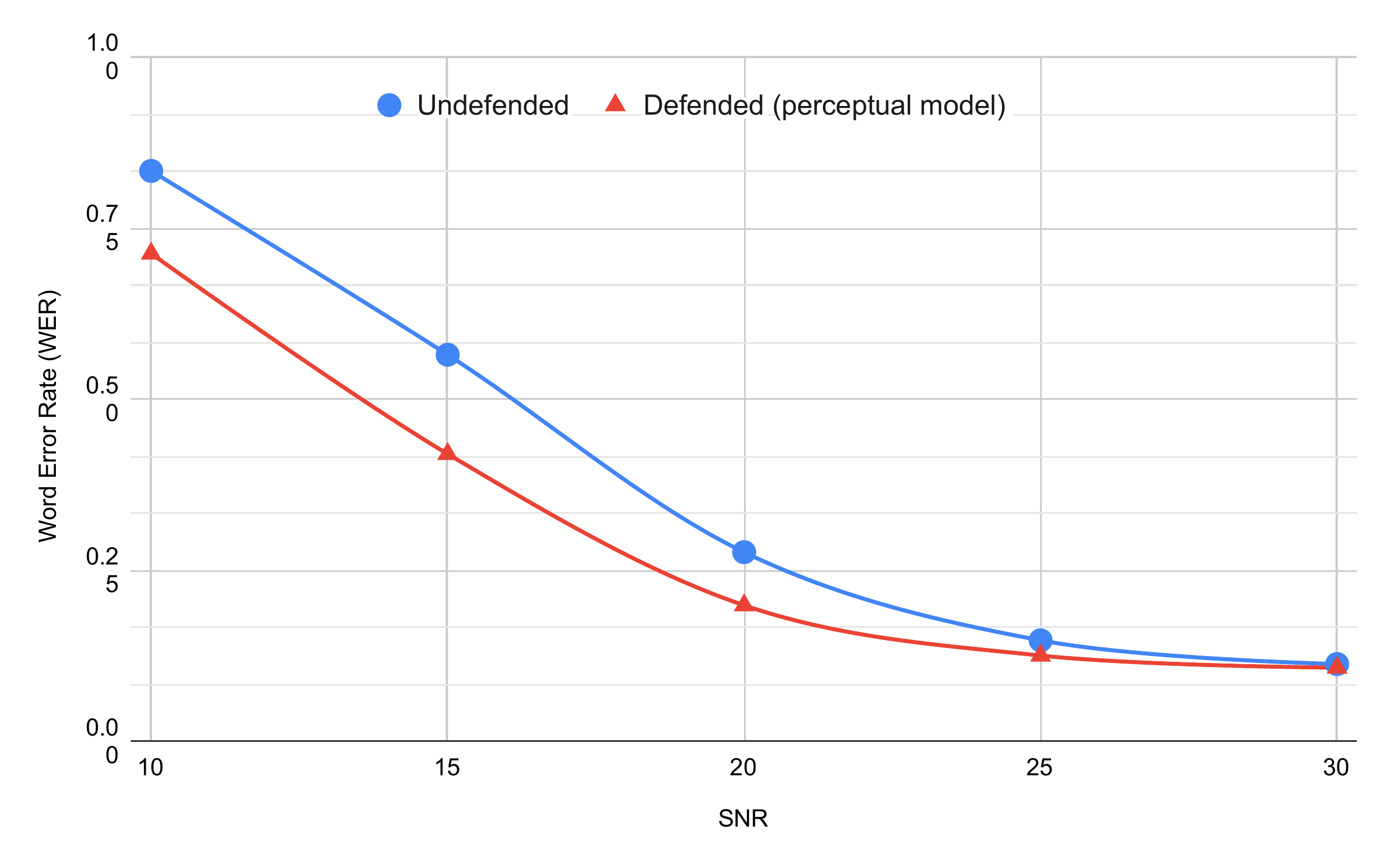}}
\caption{Word error rate (WER) for the undefended and defended models against the Kenansville attack at different adversarial SNR levels}
\label{fig:KenWER}
\end{figure}

Fig.~\ref{fig:KenWER} shows the word-error rates (WER) for the defended and undefended models against the Kenansville attack across a variety of attack SNR values. The addition of the pre-processor defense substantially improves the performance particularly for the more aggressive attacks.  Table \ref{table:2} shows the WER metrics for the MP3 defense, and the pre-trained denoiser without the perceptual loss function. From table \ref{table:2} it is clear that the MP3 defense was not effective against attacks at all SNRs and in fact performs worse than even the undefended model. The pre-trained denoiser provides $6\%$ improvement in WER over the undefended model on average. Although the pre-trained denoiser performs well, the addition of the perceptual loss improved performance on the $15$-$30$ dB attacks by an average of $1\%$ across all the SNRs. We observed that our pre-processor defense substantially improved the ASR model's performance on adversarial samples without significantly compromising the benign accuracy. 
\par

Fig.~\ref{fig:spectrogram} shows the spectrograms of the benign, adversarial attack (Kenansville) on the benign at 20dB SNR, the pre-trained denoised speech signal and the proposed perceptual loss based denoised speech signal respectively. The adversarial attack distorts the benign speech signal and the pre-trained denoiser tries to counter the attack and the proposed denoiser further tries to make the attacked signal more close to the benign speech signal.

\begin{figure}[htb]
\centerline{\includegraphics[width=8.5cm]{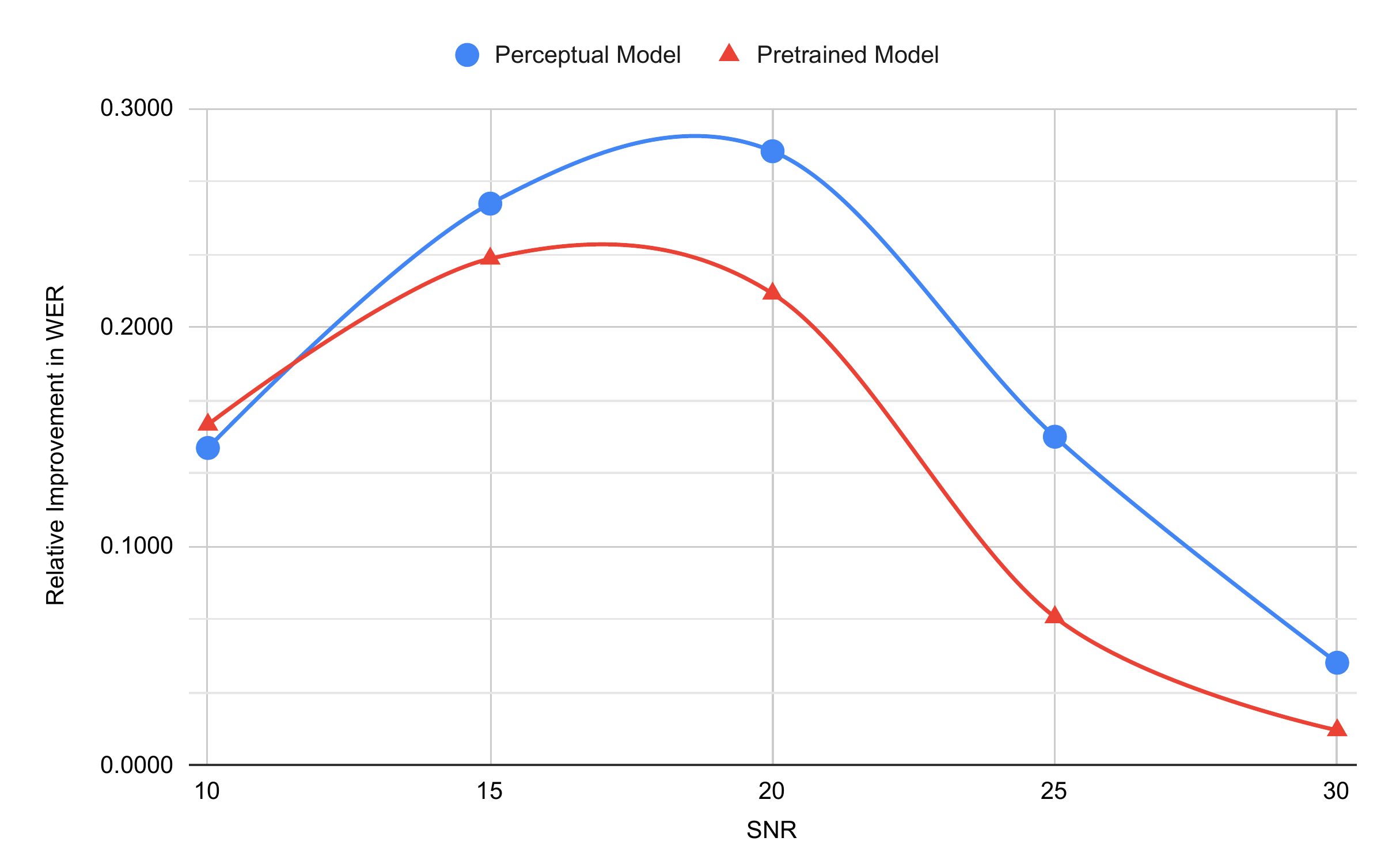}}
\caption{Relative improvement in word error rate (WER) over the undefended model for the Kenansville attack. The results for both the pre-trained denoiser and perceptual loss based denoiser training with the perceptual loss are shown.}
\label{fig:KenRelWER}
\end{figure}

On the most aggressive attack ($10$ dB), the pre-trained model actually outperformed the perceptual denoiser. In practice, however, such a low SNR is highly unlikely as an attack scenario, since it would almost certainly be detected by a human listener, which beats the purpose of an imperceptible attack. The relative improvements in WER over the undefended model are shown in Fig.~\ref{fig:KenRelWER}, which further highlights the advantages gained through the addition of the perceptual loss function. 
\par 

As shown in Table~\ref{table:2}, the proposed defense had no substantial impact on the baseline ASR performance on benign samples, unlike previously proposed methods such as adversarial training \cite{kenansville}. Once again, the MP3 defense performed poorly, markedly increasing the benign WER. 

\section{Conclusion}
\label{sec:majhead}

In this paper we presented a novel pre-processor defense against adversarial attacks on ASR systems. This defense consists of a deep neural network denoiser trained on a perceptual loss function. It is independent of the downstream ASR model, and thus can easily be deployed alongside existing ASR systems. When tested against the Kenansville attack, it produced an average improvement in WER of approximatley $6\%$ over the undefended model. Furthermore, the addition of the perceptual loss function to the denoiser training improved the the adversarial WER over the model trained with the L1 and STFT losses alone. Finally, unlike other defense strategies it did not degrade the ASR's baseline performance on benign speech samples.
\par
In the future, we would like to explore methods to further improve the denoiser's adversarial robustness. For example, it may be advantageous to train the denoising model on audio samples with attack-specific perturbations instead of additive Gaussian noise. We would also like to evaluate the defense against a wider set of adversarial attacks, specifically white-box strategies. Since these attacks have specific knowledge about an ASR model, they are generally more challenging to defend against. It remains to be seen whether or not, and under what conditions, the proposed defense would be able to successfully counteract them.

\bibliographystyle{IEEEbib}
\bibliography{strings,refs}

\end{document}